\documentclass[aps,twocolumn,showpacs,superscriptaddress]{revtex4}

\usepackage[dvips]{graphicx}

\begin{document}

\bibliographystyle{apsrev}


\title{Superconducting transition temperatures and coherence length in 
   non--\lowercase{$s$}-wave pairing materials correlated with spin-fluctuation
   mediated interaction}

\author{G. G. N. Angilella}
\affiliation{Dipartimento di Fisica e Astronomia, Universit\`a di
   Catania, and Istituto Nazionale per la Fisica della Materia, UdR
   di Catania,\\ Corso Italia, 57, I-95129 Catania, Italy}

\author{N. H. March}
\affiliation{Oxford University, Oxford, England}
\affiliation{Department of Physics, University of Antwerp (RUCA),
   Antwerp, Belgium}

\author{R. Pucci}
\affiliation{Dipartimento di Fisica e Astronomia, Universit\`a di
   Catania, and Istituto Nazionale per la Fisica della Materia, UdR
   di Catania,\\ Corso Italia, 57, I-95129 Catania, Italy}

\date{\today}

\begin{abstract}
Following earlier work on electron or hole liquids flowing through
   assemblies with magnetic fluctuations, we have recently exposed a
   marked correlation of the superconducting temperature $T_c$, for
   non--$s$-wave pairing materials, with coherence length $\xi$ and
   effective mass $m^\ast$.
The very recent study of Abanov et al. [Europhys. Lett. {\bf
   54}, 488 (2001)] and the prior investigation of Monthoux and
   Lonzarich [Phys. Rev. B {\bf 59}, 14598 (1999)] have each focussed
   on the concept of a spin-fluctuation temperature $T_{\mathrm{sf}}$, 
   which again is intimately related to $T_c$.
For the $d$-wave pairing via antiferromagnetic spin fluctuations in
   the cuprates, these studies are brought into close contact with our 
   own work, and the result is that $k_{\mathrm{B}} T_{\mathrm{sf}}
   \sim \hbar^2 / m^\ast \xi^2$.
This demonstrates that $\xi$ is also determined by such
   antiferromagnetic spin-fluctuation mediated pair interaction. 
The coherence length in units of the lattice spacing is then
   essentially given in the cuprates as the square root of the ratio
   of two characteristic energies, namely: the kinetic energy of
   localization of a charge carrier of mass $m^\ast$ in a specified
   magnetic correlation length to the hopping energy.
The quasi-2D ruthenate Sr$_2$RuO$_4$, with $T_c \sim 1.3$~K, has
   $p$-wave spin-triplet pairing and so is also briefly discussed
   here.
\end{abstract}


\maketitle

\section{Introduction}
\label{sec:intro}

In early work, Egorov and March \cite{Egorov:94} discussed electron or 
   hole liquids flowing through assemblies with antiferromagnetic spin 
   fluctuations, and proposed a correlation between the in-plane electrical
   resistivity $\rho_{ab}$ and the nuclear-spin lattice relaxation time
   $T_1$ of the form
\begin{equation}
\rho_{ab} T_1 \propto T.
\label{eq:Egorov}
\end{equation}
This relationship has been tested on the underdoped high-$T_c$ cuprate
   YBa$_2$Cu$_4$O$_8$ with $T_1$ extracted from $^{63}$Cu NMR data, and is
   appropriate somewhat above the 
   superconducting transition temperature (see Fig.~A.7.5.1 on p.~355 of
   Ref.~\onlinecite{March:96a}, and Ref.~\onlinecite{Angilella:00a}).
The correlation written in Eq.~(\ref{eq:Egorov}) between the observables 
   $\rho_{ab}$ and $T_1$ arose \cite{Egorov:94} by eliminating the magnetic 
   susceptibility $\chi({\bf Q})$, at the antiferromagnetic wave
   vector ${\bf Q}$ [equal to $(\pi/a,\pi/a)$ in the treatment below,
   $a$ being the lattice spacing], with $\rho_{ab}$ and $(TT_1 )^{-1}$ both 
   proportional to $\chi({\bf Q})$, as shown by Kohno and Yamada
   \cite{Kohno:91}.

More recently, the present authors \cite{Angilella:00b} have
   demonstrated that $T_c$ for non--$s$-wave pairing superconductors,
   and in particular for heavy Fermion materials and high-$T_c$
   cuprates, correlated with coherence length and effective mass.
The very recent and apparently quite different studies of Abanov et
   al. \cite{Abanov:01}, and prior to that, of Monthoux and Lonzarich
   \cite{Monthoux:99}, are here brought into close contact with our
   earlier work \cite{Angilella:00b}.

The outline of this Brief Report is then as follows.
In Sec.~\ref{sec:pd}, we briefly summarize the essential input in the
   treatments of spin-fluctuation mediated pairing in
   Refs.~\onlinecite{Abanov:01} and \onlinecite{Monthoux:99}.
Anticipating that, we stress at the outset that the common feature in
   Refs.~\onlinecite{Abanov:01} and \onlinecite{Monthoux:99} is a
   characteristic thermal energy $k_{\mathrm{B}} T_{\mathrm{sf}}$
   associated with a spin-fluctuation temperature $T_{\mathrm{sf}}$.
Sec.~\ref{sec:TsfXi} connects the work on $T_{\mathrm{sf}}$ in
   Refs.~\onlinecite{Abanov:01} and \onlinecite{Monthoux:99} with our
   own studies on coherence length.
A summary is then given in Sec.~\ref{sec:summary}, with some possible
   directions for future work.

\section{Spin-fluctuation temperature related to $T_c$ in
   \lowercase{$p$}- and \lowercase{$d$}-wave superconductivity in
   quasi-2D metals}
\label{sec:pd}

Essential input into both Refs.~\onlinecite{Abanov:01} and
   \onlinecite{Monthoux:99} is a form of the retarded generalized
   magnetic susceptibility $\chi({\bf q},\omega)$.
Quite specifically, in Ref.~\onlinecite{Monthoux:99} the
   phenomenological form
\begin{equation}
\chi({\bf q},\omega) = \frac{\chi_0 \kappa_0^2}{\kappa^2 + \hat{q}^2 -
   i[\omega/\eta(\hat{q})]}
\label{eq:chiML}
\end{equation}
is assumed.
Here, $\kappa^{-1}$ and $\kappa_0^{-1}$ are correlation lengths in
   units of the lattice constant $a$, with and without strong magnetic 
   correlations, respectively.

Monthoux and Lonzarich \cite{Monthoux:99} then adopt a tight-binding
   form for the quasiparticle dispersion relation, and subsequently
   define the quantities $\hat{q}_\pm^2$ by
\begin{equation}
\hat{q}_\pm^2 = 4 \pm 2[\cos(q_x a) + \cos(q_y a)].
\end{equation}
In the case of ferromagnetic correlations, typified by the ruthenate
   Sr$_2$RuO$_4$ with a low $T_c \sim 1.3$~K \cite{Maeno:94}, the
   parameters $\hat{q}^2$ and $\eta(\hat{q})$ entering $\chi({\bf
   q},\omega)$ in Eq.~(\ref{eq:chiML}) are defined as
\begin{equation}
\hat{q}^2 = \hat{q}^2_-
\end{equation}
and
\begin{equation}
\eta(\hat{q}) = T_{\mathrm{sf}} \hat{q}_- ,
\end{equation}
where $T_{\mathrm{sf}}$ is the spin-fluctuation temperature already
   referred to in Sec.~\ref{sec:intro}.


Here, we should caution that the use of a single-band dispersion relation
   [Eq.~(3) in Ref.~\onlinecite{Monthoux:99}], while appropriate for
   most high-$T_c$ cuprates, is clearly a rather crude
   approximation for Sr$_2$RuO$_4$.
Indeed, this compound is known to be characterized by
   a three-fold band, arising from the hybridization of the ruthenium
   $4d_{ij}$ atomic orbitals ($ij = xy$, $xz$, $yz$) in the $t_{2g}$
   subshell.
As a consequence, the Fermi surface (FS) is composed of a set of three
   disconnected barrel-like sheets, almost dispersionless in the
   direction orthogonal to the SrO$_2$ layers \cite{Mackenzie:96}.
This led to the proposal of an orbital-dependent form of
   superconductivity \cite{Agterberg:97}, where the role of the
   multiband nature of Sr$_2$RuO$_4$ in stabilizing a
   $p$-wave triplet order parameter against a $d$-wave singlet
   alternative has been also emphasized \cite{Mazin:99}.
However, we believe that the main conclusions of Monthoux and Lonzarich
   \cite{Monthoux:99} would be unaffected by multiband
   effects, at least qualitatively.
Moreover, the restriction to the main band of Sr$_2$RuO$_4$ [Eq.~(3)
   in Ref.~\onlinecite{Monthoux:99}] helps treating the quasi-2D
   character of the single-particle dispersion relation of this
   layered perovskite on the same footing as for the cuprates
   \cite{note:1}.


Monthoux and Lonzarich \cite{Monthoux:99} also investigate
   antiferromagnetic correlations as in the $d$-wave paired cuprates,
   in which case the above parameters have the form
\begin{equation}
\hat{q}^2 = \hat{q}^2_+
\end{equation}
and
\begin{equation}
\eta(\hat{q}) = T_{\mathrm{sf}} \hat{q}_- .
\end{equation}
The final input we need to refer to here is a coupling parameter $g^2$ 
   in the quasiparticle self-energy $\Sigma({\bf q},\omega)$,
   involving of course [see Ref.~\onlinecite{Monthoux:99}, Eqs.~(11)--(13)]
   summations over wave vectors and Matsubara frequencies of
   $\chi({\bf q},\omega)$.

The mean-field Eliashberg equations for nearly ferromagnetic and
   nearly antiferromagnetic metals with a single 2D Fermi surface were 
   then solved numerically in Ref.~\onlinecite{Monthoux:99}, to obtain 
   the ratio of critical temperature $T_c$ to spin fluctuation
   temperature, $T_{\mathrm{sf}}$, essentially as a function of
   coupling strength $g^2$ for different values of the inverse
   correlation length $\kappa$.
This was done both for $p$-wave triplet and $d$-wave singlet pairing.

The major predictions of Refs.~\onlinecite{Abanov:01} and
   \onlinecite{Monthoux:99} were in accord that at strong coupling,
   $T_c / T_{\mathrm{sf}}$ exhibits saturation.
For a physically reasonable range of values of $\kappa^{-1}$, the
   quantitative results of Ref.~\onlinecite{Monthoux:99} were:
   \emph{(i)} For $p$-wave triplet pairing, $T_c / T_{\mathrm{sf}}$
   saturates at a value of $1/30$, and \emph{(ii)} For $d$-wave
   singlet pairing, $T_c / T_{\mathrm{sf}}$ has a saturation value of
   $1/2$.
This then is the point at which to make contact between these findings 
   of Refs.~\onlinecite{Abanov:01} and \onlinecite{Monthoux:99} and
   our own study \cite{Angilella:00b}.

\section{Spin-fluctuation temperature $T_{\mathrm{sf}}$ and
   correlation length $\xi$ in non--\lowercase{$s$}-wave pairing
   superconductors: especially high-$T_c$ cuprates}
\label{sec:TsfXi}

In Ref.~\onlinecite{Angilella:00b}, we exposed a relationship, for
   both heavy Fermion materials and for high-$T_c$ cuprates, between
   the thermal energy $k_{\mathrm{B}} T_c$ and another characteristic
   energy, $\epsilon_c$ say, for such non--$s$-wave superconductors,
   where $\epsilon_c$ was defined by
\begin{equation}
\epsilon_c = \frac{\hbar^2}{m^\ast \xi^2} ,
\label{eq:epsc}
\end{equation}
$\xi$ being the coherence length and $m^\ast$ the effective mass.
We noted \cite{Angilella:00b} that Uemura et al. \cite{Uemura:91} had
   already clearly recognized that $m^\ast$ should enter inversely in
   determining the scale of $k_{\mathrm{B}} T_c$.

Since Monthoux and Lonzarich \cite{Monthoux:99} made their most
   extensive numerical investigations for the $p$-wave triplet
   pairing, let us take first the ruthenate Sr$_2$RuO$_4$, discussed
   at some length in Ref.~\onlinecite{Monthoux:99}.
From Fig.~2a of Ref.~\onlinecite{Monthoux:99}, for example, provided
   $\kappa^2$ is in the (physically reasonable) range from 0.25 to 1.0 
   and their quantity $g^2 \chi_0 /t$, with $t$ the hopping energy, is 
   in the physical range 10--20, then $T_c /T_{\mathrm{sf}}$ is in the 
   range 0.02--0.03, which yields
   $T_{\mathrm{sf}} \simeq 50$~K for Sr$_2$RuO$_4$, compared with a
   `saturation' value of 40~K for this material with $T_c \sim 1.3$~K.
Abanov et al. \cite{Abanov:01} refer to values of $T_{\mathrm{sf}}
   \sim 100$~K, so there is semiquantitative accord.

Turning to the high-$T_c$ cuprates, the present authors
   \cite{Angilella:00b} have pointed out that in marked contrast to
   the heavy Fermion materials they also considered, $k_{\mathrm{B}}
   T_c \sim \hbar^2 /m^\ast \xi^2$ and from the saturation limit
   \emph{(ii)} in Sec.~\ref{sec:pd}, taken again from
   Ref.~\onlinecite{Monthoux:99},
\begin{equation}
k_{\mathrm{B}} T_c \sim \frac{1}{2} k_{\mathrm{B}} T_{\mathrm{sf}}
\end{equation}
for the $d$-wave pairing high-$T_c$ cuprates.
Thus, one has as a consequence the order of magnitude result
\begin{equation}
k_{\mathrm{B}} T_{\mathrm{sf}} \sim \frac{2\hbar^2}{m^\ast \xi^2},
\label{eq:oom}
\end{equation}
and the coherence length $\xi$ of the high-$T_c$ cuprates is plainly
   determined by the antiferromagnetic spin-fluctuation mediated
   pairing, via a temperature $T_{\mathrm{sf}} \sim 2 T_c$.

\subsection{Physical interpretation of the coherence length $\xi$
   resulting from \lowercase{$d$}-wave singlet pairing
   spin-fluctuation interaction in the high-$T_c$ cuprates}
\label{ssec:interpretation}

We return to our earlier result \cite{Angilella:00b} obtained from
   experimental data for the high-$T_c$ cuprates that $k_{\mathrm{B}}
   T_c \sim\epsilon_c$, with $\epsilon_c$ given by Eq.~(\ref{eq:epsc}).
We now add to this empirical correlation a further experimental
   consequence used by Monthoux and Lonzarich \cite{Monthoux:99},
   namely that the product of the thermal energy $k_{\mathrm{B}}
   T_{\mathrm{sf}}$ associated with the spin-fluctuation temperature
   \cite{Abanov:01,Monthoux:99} $T_{\mathrm{sf}}$ with $\kappa_0^2$,
   the inverse magnetic correlation length squared without strong
   magnetic correlations, is constant, i.e.
\begin{equation}
k_{\mathrm{B}} T_{\mathrm{sf}} \kappa_0^2 = \mathrm{const}.
\end{equation}
Adopting the value in their Table~5, the constant value turns out to
   be $\sim 8t$.

Returning to the strong coupling limit to gain further insight into
   the factors determining the coherence length $\xi$, we have
\begin{equation}
k_{\mathrm{B}} T_c \sim \frac{1}{2} k_{\mathrm{B}} T_{\mathrm{sf}}
   \sim \frac{4t}{\kappa_0^2} .
\end{equation}
Putting $\kappa_0^2 = a^2 / \ell_{m0}^2$, where $\ell_{m0}$ is the
   (antiferro-) magnetic correlation length in the high-$T_c$
   cuprates, we find almost immediately
\begin{equation}
\xi \sim \frac{a}{2} \left( \frac{\hbar^2}{m^\ast \ell_{m0}^2}
   \right)^{1/2} \frac{1}{t^{1/2}}.
\label{eq:xi}
\end{equation}

Physically, Eq.~(\ref{eq:xi}) shows that, in units of the lattice
   spacing $a$, the magnitude of the coherence length is determined by 
   the square root of the ratio of two further characteristic
   energies.
The first of these is the kinetic energy of localization of a carrier
   of mass $m^\ast$ within the magnetic correlation length $\ell_{m0}$ 
   determined however in the absence of strong magnetic correlations.
The second energy is $t$, the magnitude of the hopping energy.

We can anticipate, using the additional (generally weaker!) variables
   $g^2 \chi_0 /t$ and $\kappa^2$ that, provided the range of $g^2
   \chi_0 /t$ is limited to the physical region 10 to 20 (see
   Ref.~\onlinecite{Monthoux:99}) and $\kappa^2$ is likewise
   restricted to the range 0.5 to 1, then Eq.~(\ref{eq:xi}) will be
   replaced, away from the strong coupling limit, by
\begin{equation}
\xi \sim \frac{a}{2} \left( \frac{\hbar^2}{m^\ast \ell_{m0}^2}
   \right)^{1/2} \frac{1}{t^{1/2}} F(g^2 \chi_0 /t ; \kappa^2),
\label{eq:xigeneral}
\end{equation}
where $F$ is a slowly varying function of its arguments, $F$ becoming
   unity for sufficiently large values of the `coupling strength' $g^2
   \chi_0 /t$, with $\kappa^2$ restricted to the range quoted above.

\section{Summary and directions for future work}
\label{sec:summary}

The achievement of the present Brief Report is to bring the studies of 
   Refs.~\onlinecite{Abanov:01} and \onlinecite{Monthoux:99}, in which 
   the superconducting transition temperature $T_c$ is connected to the
   spin-fluctuation temperature $T_{\mathrm{sf}}$, into direct contact 
   with our work relating $k_{\mathrm{B}} T_c$ to the characteristic
   energy $\hbar^2 /m^\ast \xi^2$ \cite{Angilella:00b}.
For a $d$-wave singlet pairing mediated by antiferromagnetic spin
   fluctuations in the cuprates, the simple, order of magnitude
   relation Eq.~(\ref{eq:oom}) follows, showing that the coherence
   length $\xi$ is determined by the interaction mediated by spin
   fluctuations.
This is expressed, more specifically, in the language of
   Ref.~\onlinecite{Monthoux:99}, in Eqs.~(\ref{eq:xi}) and
   (\ref{eq:xigeneral}).

\begin{figure}
\centering
\includegraphics[height=\columnwidth,angle=-90]{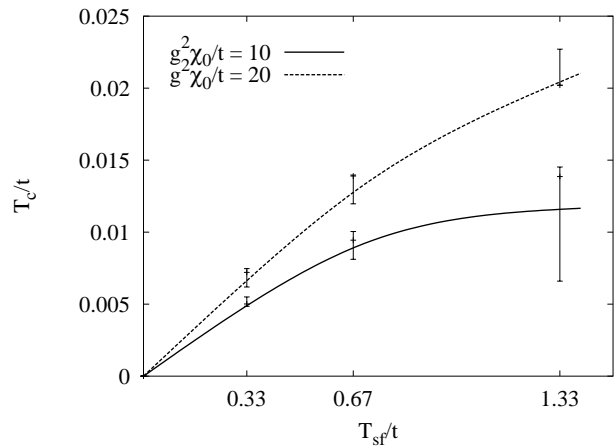}
\caption{%
Shows plot of $T_c/t$ vs $T_{\mathrm{sf}}/t$ for $p$-wave spin-triplet
   pairing.
This has been constructed by combining numerical data from Figs.~2--4
   of Ref.~\protect\onlinecite{Monthoux:99}, assuming for the
   parameters the physical range $10\leq g^2 \chi_0 /t\leq 20$ and
   $0.25\leq\kappa^2\leq1$.
Points corresponding to the same value of $T_{\mathrm{sf}}/t$ and of $g^2 
   \chi_0 /t$, but to different values of $\kappa^2$, have been arranged 
   as vertical bars.
The two curves shown are guides to the eye through the choices of
   `coupling strength' $g^2 \chi_0 /t$ of 10 and 20, and have been
   extrapolated through the origin.
}
\label{fig:Monthoux}
\end{figure}

However, the situation regarding the relation of $T_{\mathrm{sf}}$ in
   the low-$T_c$ ruthenate Sr$_2$RuO$_4$ to the coherence length $\xi$ 
   is much less clear presently than in the high-$T_c$ cuprates.
This may be because of a competition between nearly ferromagnetic
   behavior and antiferromagnetic spin fluctuations \cite{Mazin:99}.
Experiments on $\chi({\bf q},\omega)$ using both neutron scattering
   and NMR on this ruthenate would be valuable for furthering
   understanding of the origins of superconductivity, and especially
   the physics of the coherence length in this material.
As for $T_{\mathrm{sf}}$ it seems to lie in the range 40--50~K.
Having referred to heavy Fermion materials in connection with
   Ref.~\onlinecite{Angilella:00b}, 
we thought it of interest to construct
   from the $p$-wave studies of Ref.~\onlinecite{Monthoux:99} a plot
   of $T_c/t$ vs $T_{\mathrm{sf}}/t$ (see Fig.~\ref{fig:Monthoux}), by
   combining data from their Figs.~2--4.
It is worth noting, though we expect the mechanisms generally to be
   different, that the shape of the present Fig.~\ref{fig:Monthoux}
   parallels that of Fig.~1 of Ref.~\onlinecite{Angilella:00b}.
However, such a comparison should be carefully considered, since some
   of the heavy Fermion materials considered in Fig.~1 of
   Ref.~\onlinecite{Angilella:00b} exhibit antiferromagnetic spin
   fluctuations, rather than ferromagnetic spin fluctuations, as
   studied by Monthoux and Lonzarich in connection with $p$-wave
   superconductors \cite{Monthoux:99}.
The major exception is UPd$_2$Al$_3$, which is known to be
   characterized by rather strong, static antiferromagnetic
   correlations in the normal state (with a significant N\'eel
   temperature of $T_{\mathrm{N}} = 14.3$~K).
Such antiferromagnetic correlations even coexist with
   superconductivity below $T_c \sim 2$~K, at variance with other
   uranium based heavy Fermion compounds
   \cite{Bernhoeft:99,Steglich:00,Naidyuk:01}.
We record, however, that this compound, with its relatively high
   $T_c$, helped Sato (Fig.~2 in Ref.~\onlinecite{Sato:99}) to
   `bridge the gap' between the low-$T_c$ heavy Fermion compounds and
   the high-$T_c$ cuprates in establishing a correlation between $T_c$
   and some magnetic ordering temperature (though related to a
   different kind of magnetic order in different compounds), much in the
   same spirit as in the present work (compare also Fig.~2 of
   Ref.~\onlinecite{Angilella:00b}).

\begin{acknowledgments}
One of us (N.H.M.) made his contribution to the present Report during
   a visit to the Physics Department, University of Catania, in the
   year 2001.
Thanks are due to the Department for the stimulating environment and
   for much hospitality.
G.G.N.A. thanks Dr. G. Sparn for stimulating discussions and
   correspondence, and acknowledges partial support from the EU
   through the FSE program. 
\end{acknowledgments}

\bibliography{a,b,c,d,e,f,g,h,i,j,k,l,m,n,o,p,q,r,s,t,u,v,w,x,y,z,zzproceedings,Angilella,notes}

\begin{thebibliography}{18}
\expandafter\ifx\csname natexlab\endcsname\relax\def\natexlab#1{#1}\fi
\expandafter\ifx\csname bibnamefont\endcsname\relax
  \def\bibnamefont#1{#1}\fi
\expandafter\ifx\csname bibfnamefont\endcsname\relax
  \def\bibfnamefont#1{#1}\fi
\expandafter\ifx\csname citenamefont\endcsname\relax
  \def\citenamefont#1{#1}\fi
\expandafter\ifx\csname url\endcsname\relax
  \def\url#1{\texttt{#1}}\fi
\expandafter\ifx\csname urlprefix\endcsname\relax\def\urlprefix{URL }\fi
\providecommand{\bibinfo}[2]{#2}
\providecommand{\eprint}[2][]{\url{#2}}

\bibitem[{\citenamefont{Egorov and March}(1994)}]{Egorov:94}
\bibinfo{author}{\bibfnamefont{S.}~\bibnamefont{Egorov}} \bibnamefont{and}
  \bibinfo{author}{\bibfnamefont{N.~H.} \bibnamefont{March}},
  \bibinfo{journal}{Phys. Chem. Liquids} \textbf{\bibinfo{volume}{28}},
  \bibinfo{pages}{141} (\bibinfo{year}{1994}).

\bibitem[{\citenamefont{March}(1996)}]{March:96a}
\bibinfo{author}{\bibfnamefont{N.~H.} \bibnamefont{March}},
  \emph{\bibinfo{title}{Electron correlation in atoms, molecules and condensed
  phases}} (\bibinfo{publisher}{Plenum Press}, \bibinfo{address}{New York},
  \bibinfo{year}{1996}), \bibinfo{note}{p. 354}.

\bibitem[{\citenamefont{Angilella
  et~al.}(2000{\natexlab{a}})\citenamefont{Angilella, March, and
  Pucci}}]{Angilella:00a}
\bibinfo{author}{\bibfnamefont{G.~G.~N.} \bibnamefont{Angilella}},
  \bibinfo{author}{\bibfnamefont{N.~H.} \bibnamefont{March}}, \bibnamefont{and}
  \bibinfo{author}{\bibfnamefont{R.}~\bibnamefont{Pucci}},
  \bibinfo{journal}{Phys. Chem. Liquids} \textbf{\bibinfo{volume}{38}},
  \bibinfo{pages}{615} (\bibinfo{year}{2000}{\natexlab{a}}).

\bibitem[{\citenamefont{Kohno and Yamada}(1991)}]{Kohno:91}
\bibinfo{author}{\bibfnamefont{H.}~\bibnamefont{Kohno}} \bibnamefont{and}
  \bibinfo{author}{\bibfnamefont{K.}~\bibnamefont{Yamada}},
  \bibinfo{journal}{Prog. Theor. Phys.} \textbf{\bibinfo{volume}{85}},
  \bibinfo{pages}{13} (\bibinfo{year}{1991}).

\bibitem[{\citenamefont{Angilella
  et~al.}(2000{\natexlab{b}})\citenamefont{Angilella, March, and
  Pucci}}]{Angilella:00b}
\bibinfo{author}{\bibfnamefont{G.~G.~N.} \bibnamefont{Angilella}},
  \bibinfo{author}{\bibfnamefont{N.~H.} \bibnamefont{March}}, \bibnamefont{and}
  \bibinfo{author}{\bibfnamefont{R.}~\bibnamefont{Pucci}},
  \bibinfo{journal}{Phys. Rev. B} \textbf{\bibinfo{volume}{62}},
  \bibinfo{pages}{13919} (\bibinfo{year}{2000}{\natexlab{b}}).

\bibitem[{\citenamefont{{Ar. Abanov} et~al.}(2001)\citenamefont{{Ar. Abanov},
  Chubukov, and Finkel'stein}}]{Abanov:01}
\bibinfo{author}{\bibnamefont{{Ar. Abanov}}},
  \bibinfo{author}{\bibfnamefont{A.~V.} \bibnamefont{Chubukov}},
  \bibnamefont{and} \bibinfo{author}{\bibfnamefont{A.~M.}
  \bibnamefont{Finkel'stein}}, \bibinfo{journal}{Europhys. Lett.}
  \textbf{\bibinfo{volume}{54}}, \bibinfo{pages}{488} (\bibinfo{year}{2001}).

\bibitem[{\citenamefont{Monthoux and Lonzarich}(1999)}]{Monthoux:99}
\bibinfo{author}{\bibfnamefont{P.}~\bibnamefont{Monthoux}} \bibnamefont{and}
  \bibinfo{author}{\bibfnamefont{G.~G.} \bibnamefont{Lonzarich}},
  \bibinfo{journal}{Phys. Rev. B} \textbf{\bibinfo{volume}{59}},
  \bibinfo{pages}{14598} (\bibinfo{year}{1999}).

\bibitem[{\citenamefont{Maeno et~al.}(1994)\citenamefont{Maeno, Hashimoto,
  Yoshida, Nishizaki, Fujita, Bednorz, and Lichtenberg}}]{Maeno:94}
\bibinfo{author}{\bibfnamefont{Y.}~\bibnamefont{Maeno}},
  \bibinfo{author}{\bibfnamefont{H.}~\bibnamefont{Hashimoto}},
  \bibinfo{author}{\bibfnamefont{K.}~\bibnamefont{Yoshida}},
  \bibinfo{author}{\bibfnamefont{S.}~\bibnamefont{Nishizaki}},
  \bibinfo{author}{\bibfnamefont{T.}~\bibnamefont{Fujita}},
  \bibinfo{author}{\bibfnamefont{J.~G.} \bibnamefont{Bednorz}},
  \bibnamefont{and}
  \bibinfo{author}{\bibfnamefont{F.}~\bibnamefont{Lichtenberg}},
  \bibinfo{journal}{Nature} \textbf{\bibinfo{volume}{372}},
  \bibinfo{pages}{532} (\bibinfo{year}{1994}).

\bibitem[{\citenamefont{Mackenzie et~al.}(1996)\citenamefont{Mackenzie, Julian,
  Diver, {McMullan}, Ray, Lonzarich, Maeno, Nishizaki, and
  Fujita}}]{Mackenzie:96}
\bibinfo{author}{\bibfnamefont{A.~P.} \bibnamefont{Mackenzie}},
  \bibinfo{author}{\bibfnamefont{S.~R.} \bibnamefont{Julian}},
  \bibinfo{author}{\bibfnamefont{A.~J.} \bibnamefont{Diver}},
  \bibinfo{author}{\bibfnamefont{G.~J.} \bibnamefont{{McMullan}}},
  \bibinfo{author}{\bibfnamefont{M.~P.} \bibnamefont{Ray}},
  \bibinfo{author}{\bibfnamefont{G.~G.} \bibnamefont{Lonzarich}},
  \bibinfo{author}{\bibfnamefont{Y.}~\bibnamefont{Maeno}},
  \bibinfo{author}{\bibfnamefont{S.}~\bibnamefont{Nishizaki}},
  \bibnamefont{and} \bibinfo{author}{\bibfnamefont{T.}~\bibnamefont{Fujita}},
  \bibinfo{journal}{Phys. Rev. Lett.} \textbf{\bibinfo{volume}{76}},
  \bibinfo{pages}{3786} (\bibinfo{year}{1996}).

\bibitem[{\citenamefont{Agterberg et~al.}(1997)\citenamefont{Agterberg, Rice,
  and Sigrist}}]{Agterberg:97}
\bibinfo{author}{\bibfnamefont{D.~F.} \bibnamefont{Agterberg}},
  \bibinfo{author}{\bibfnamefont{T.~M.} \bibnamefont{Rice}}, \bibnamefont{and}
  \bibinfo{author}{\bibfnamefont{M.}~\bibnamefont{Sigrist}},
  \bibinfo{journal}{Phys. Rev. Lett.} \textbf{\bibinfo{volume}{78}},
  \bibinfo{pages}{3374} (\bibinfo{year}{1997}).

\bibitem[{\citenamefont{Mazin and Singh}(1999)}]{Mazin:99}
\bibinfo{author}{\bibfnamefont{I.~I.} \bibnamefont{Mazin}} \bibnamefont{and}
  \bibinfo{author}{\bibfnamefont{D.~J.} \bibnamefont{Singh}},
  \bibinfo{journal}{Phys. Rev. Lett.} \textbf{\bibinfo{volume}{82}},
  \bibinfo{pages}{4324} (\bibinfo{year}{1999}).

\bibitem[{not()}]{note:1}
\bibinfo{note}{The role of reduced dimensionality in enhancing magnetically
  mediated superconductivity, as emphasized by Monthoux and Lonzarich
  \cite{Monthoux:99}, has been recently confirmed experimentally in the layered
  heavy Fermion compound CeCoIn$_5$ \cite{Nicklas:01}. There, pressure $P$ is
  employed to tune the anisotropy of this material. Qualitative agreement is
  found between Monthoux and Lonzarich's predictions and the experimental trend
  of $T_c = T_c (P)$, with $P$ roughly identified with parameter $\kappa^2$ in
  Ref.~\onlinecite{Monthoux:99}.}

\bibitem[{\citenamefont{Uemura et~al.}(1991)\citenamefont{Uemura, Le, Luke,
  Sternlieb, Wu, Brewer, Riseman, Seaman, Maple, Ishikawa et~al.}}]{Uemura:91}
\bibinfo{author}{\bibfnamefont{Y.~J.} \bibnamefont{Uemura}},
  \bibinfo{author}{\bibfnamefont{L.~P.} \bibnamefont{Le}},
  \bibinfo{author}{\bibfnamefont{G.~M.} \bibnamefont{Luke}},
  \bibinfo{author}{\bibfnamefont{B.~J.} \bibnamefont{Sternlieb}},
  \bibinfo{author}{\bibfnamefont{W.~D.} \bibnamefont{Wu}},
  \bibinfo{author}{\bibfnamefont{J.~H.} \bibnamefont{Brewer}},
  \bibinfo{author}{\bibfnamefont{T.~M.} \bibnamefont{Riseman}},
  \bibinfo{author}{\bibfnamefont{C.~L.} \bibnamefont{Seaman}},
  \bibinfo{author}{\bibfnamefont{M.~B.} \bibnamefont{Maple}},
  \bibinfo{author}{\bibfnamefont{M.}~\bibnamefont{Ishikawa}},
  \bibnamefont{et~al.}, \bibinfo{journal}{Phys. Rev. Lett.}
  \textbf{\bibinfo{volume}{66}}, \bibinfo{pages}{2665} (\bibinfo{year}{1991}).

\bibitem[{\citenamefont{Bernhoeft et~al.}(1999)\citenamefont{Bernhoeft,
  Roessli, Sato, Aso, Hiess, Lander, Endoh, and Komatsubara}}]{Bernhoeft:99}
\bibinfo{author}{\bibfnamefont{N.}~\bibnamefont{Bernhoeft}},
  \bibinfo{author}{\bibfnamefont{B.}~\bibnamefont{Roessli}},
  \bibinfo{author}{\bibfnamefont{N.}~\bibnamefont{Sato}},
  \bibinfo{author}{\bibfnamefont{N.}~\bibnamefont{Aso}},
  \bibinfo{author}{\bibfnamefont{A.}~\bibnamefont{Hiess}},
  \bibinfo{author}{\bibfnamefont{G.~H.} \bibnamefont{Lander}},
  \bibinfo{author}{\bibfnamefont{Y.}~\bibnamefont{Endoh}}, \bibnamefont{and}
  \bibinfo{author}{\bibfnamefont{T.}~\bibnamefont{Komatsubara}},
  \bibinfo{journal}{Physica B} \textbf{\bibinfo{volume}{259-261}},
  \bibinfo{pages}{614} (\bibinfo{year}{1999}).

\bibitem[{\citenamefont{Steglich et~al.}(2000)\citenamefont{Steglich, Sato,
  Tayama, L\"uhmann, Langhammer, Gegenwart, Hinze, Geibel, Lang, Sparn
  et~al.}}]{Steglich:00}
\bibinfo{author}{\bibfnamefont{F.}~\bibnamefont{Steglich}},
  \bibinfo{author}{\bibfnamefont{N.}~\bibnamefont{Sato}},
  \bibinfo{author}{\bibfnamefont{T.}~\bibnamefont{Tayama}},
  \bibinfo{author}{\bibfnamefont{T.}~\bibnamefont{L\"uhmann}},
  \bibinfo{author}{\bibfnamefont{C.}~\bibnamefont{Langhammer}},
  \bibinfo{author}{\bibfnamefont{P.}~\bibnamefont{Gegenwart}},
  \bibinfo{author}{\bibfnamefont{P.}~\bibnamefont{Hinze}},
  \bibinfo{author}{\bibfnamefont{C.}~\bibnamefont{Geibel}},
  \bibinfo{author}{\bibfnamefont{M.}~\bibnamefont{Lang}},
  \bibinfo{author}{\bibfnamefont{G.}~\bibnamefont{Sparn}},
  \bibnamefont{et~al.}, \bibinfo{journal}{Physica C}
  \textbf{\bibinfo{volume}{341-348}}, \bibinfo{pages}{691}
  (\bibinfo{year}{2000}).

\bibitem[{\citenamefont{{Yu. G. Naidyuk} et~al.}(2001)\citenamefont{{Yu. G.
  Naidyuk}, Kvitnitskaya, Jansen, Geibel, Menovsky, and Wyder}}]{Naidyuk:01}
\bibinfo{author}{\bibnamefont{{Yu. G. Naidyuk}}},
  \bibinfo{author}{\bibfnamefont{O.~E.} \bibnamefont{Kvitnitskaya}},
  \bibinfo{author}{\bibfnamefont{A.~G.~M.} \bibnamefont{Jansen}},
  \bibinfo{author}{\bibfnamefont{C.}~\bibnamefont{Geibel}},
  \bibinfo{author}{\bibfnamefont{A.~A.} \bibnamefont{Menovsky}},
  \bibnamefont{and} \bibinfo{author}{\bibfnamefont{P.}~\bibnamefont{Wyder}},
  \bibinfo{journal}{Fiz. Nizk. Temp. (Low Temp. Phys.)}
  \textbf{\bibinfo{volume}{27}}, \bibinfo{pages}{668} (\bibinfo{year}{2001}),
  \bibinfo{note}{preprint {\tt cond-mat/0101062}}.

\bibitem[{\citenamefont{Sato}(1999)}]{Sato:99}
\bibinfo{author}{\bibfnamefont{N.}~\bibnamefont{Sato}},
  \bibinfo{journal}{Physica B} \textbf{\bibinfo{volume}{259-261}},
  \bibinfo{pages}{634} (\bibinfo{year}{1999}).

\bibitem[{\citenamefont{Nicklas et~al.}(2001)\citenamefont{Nicklas, Borth,
  Lengyel, Pagliuso, Sarrao, Sidorov, Sparn, Steglich, and
  Thomson}}]{Nicklas:01}
\bibinfo{author}{\bibfnamefont{M.}~\bibnamefont{Nicklas}},
  \bibinfo{author}{\bibfnamefont{R.}~\bibnamefont{Borth}},
  \bibinfo{author}{\bibfnamefont{E.}~\bibnamefont{Lengyel}},
  \bibinfo{author}{\bibfnamefont{P.~G.} \bibnamefont{Pagliuso}},
  \bibinfo{author}{\bibfnamefont{J.~L.} \bibnamefont{Sarrao}},
  \bibinfo{author}{\bibfnamefont{V.~A.} \bibnamefont{Sidorov}},
  \bibinfo{author}{\bibfnamefont{G.}~\bibnamefont{Sparn}},
  \bibinfo{author}{\bibfnamefont{F.}~\bibnamefont{Steglich}}, \bibnamefont{and}
  \bibinfo{author}{\bibfnamefont{J.~D.} \bibnamefont{Thomson}},
  \bibinfo{journal}{J. Phys.: Condens. Matter} \textbf{\bibinfo{volume}{...}},
  \bibinfo{pages}{...} (\bibinfo{year}{2001}), \bibinfo{note}{preprint {\tt
  cond-mat/0108319}}.

\end{thebibliography}

\end{document}